\documentclass[a4paper,12pt]{article}
\usepackage{amsmath}
\usepackage{amssymb}
\usepackage{textcomp}
\usepackage{hyperref}
\usepackage{cite}
\title{Superalgebra Doubling in $AdS_3$ Superstrings }
\author{Abbas Ali\footnote{Email : aali.ph@gmail.com} \\
	Physics Department, Aligarh Muslim University,\\ Aligarh, India}
\date{}
\begin{document}

\maketitle

\begin{abstract}
We propose a solution to the mystery of superalgebra doubling found by Giveon-Kutasov-Seiberg in their construction of worldsheet approach to spacetime superconformal symmetry  for $AdS_3$ superstrings in the context of $AdS_3$/$CFT_2$  Correspondence.
\end{abstract}
\section{Introduction}

$AdS_3/CFT_2$ is one of the two most intensely studied cases of $AdS/CFT$ correspondence. This   is formulated as the $1+1$ dimensional CFT describing the Higgs branch of the D1-D5 system on $M^{4} $= $K3$ or $T^{4}$ being dual to Type IIB string theory on $(AdS_3 \times S^{3})_{Q_1Q_5} \times M^{4}(Q)$ \cite{Maldacena:1997re,Witten:1998qj,Gubser:1998bc}. This naturally generalizes to $M^{4} = S^{3} \times S^{1}$. In this regard 
Giveon-Kutasov-Seiberg (GKS) poineered the worldsheet approach to the spacetime symmetry algebra for $AdS_3$ strings and superstrings
\cite{Giveon:1998ns}.
We owe our major insights to this approach in case of $AdS_3/CFT_2$ \cite{Elitzur:1998mm,Ali:2000we}.

At the moment public attention has gyrated toward the entanglement aspects of this system but several unresolved problems are gawking at us in the core formulation of $AdS_3/CFT_2$ correspondence. Some of these are as follows.

We know the brane configurations, namely the $D1/D5$ system, that go to the  $AdS_3 \times S^3 \times K3$ and $AdS_3 \times S^3 \times T^4$ geometry in the near horizon limit. But we do not know the brane configuration that will go to the   $AdS_3 \times S^3 \times S^3 \times S^1$ geometry in the near horizon limit.

Similarly the CFT dual to superstrings moving on $AdS_3 \times S^3 \times M^4$ with $M^4=K3$ or $T^4$ is the symmetric orbifold theory $Sym^{N}(M^4)$ but the proposed dual in case of $AdS_3 \times S^3 \times S^3 \times S^1$, namely $Sym^N(S^3 \times S^1)$, is known with some conditions only     after a long series of investigations  \cite{Gukov:2004ym,Tong:2014yna,Gaberdiel:2010pz,Eberhardt:2017pty,Eberhardt:2017fsi}.  

In this story Gukov-Martinec-Moore-Strominger \cite{Gukov:2004ym} explored the possibility of a free field realization of the large $N=4$ superconformal symmetry that is more complete than the conventionally used free field realization by Sevrin-Troost-van Proeyen \cite{Sevrin:1988ab} so as to reach the holographic dual to superstrings moving on  $AdS_3 \times S^3 \times S^3 \times S^1$. A  non-commutativity did not allow  this to succeed. As a matter of fact corresponding complete free field realization is already known for a long time \cite{Ali:2000zu, Ali:2003aa,Ivanov:1988rt}. To the best of our knowledge this has not been put to good use for finding either the brane configuration that will go to $AdS_3 \times S^3 \times S^3 \times S^1$ geometry in the near horizon limit or the conformal field theory dual to superstrings moving on $AdS_3 \times S^3 \times S^3 \times S^1$ background.

Apart from D1-D5 system the study of $AdS_3$ superstrings is also closely related to the study of the three dimensional BTZ black hole \cite{Banados:1992wn}. While dealing with the corresponding issue in $AdS_3$ the stringy generalization of BTZ in Ref.\cite{Ali:1992mj} is also mostly ignored. As a result there are a number of issues that have to be addressed in this area.

The Penrose limit of $AdS_3 \times S^3 \times S^3 \times S^1$ has been studied in \cite{Sommovigo:2003kd,Dei:2018yth} but sufficient attention has not been paid to its connection to $AdS_3 \times S^3 \times T^4$ superstrings  \cite{Ali:2000we}.

It is well known that the superconformal symmetry relevant for the $AdS_3 \times S^3 \times K3$ and $AdS_3 \times S^3 \times S^3 \times S^1$ superstrings are the small and large N=4 superconfromal symmetries. But this remains largely unknown that the superconformal symmetry relevant for superstrings moving on  $AdS_3 \times S^3 \times T^4$ background is the middle N=4 superalgebra \cite{Ali:1993sd}.

D1-D5 system is intrinsic to the formulation of $AdS_3/CFT_2$ correspondence. In Ref.\cite{Seiberg:1999xz} Seiberg-Witten  analyzed in detail the marginal stability of this system. In absence of B-field and $\theta$-term the D1-D5 system has an instability. The branes can separate at no cost of energy. This results in a singularity of the corresponding conformal field theory. We do not know the structure of the corresponding singularity. 

There is an additional issue concerning the analysis of the spectrum of the chiral primary states in the D1-D5 system. The analysis of GKS found that the chiral operators in the boundary conformal field theory have $\Delta \le (Q_5-2)/2$. The same spectrum analyzed in Ref.\cite{Strominger:1996sh} has the bound $\Delta \le Q_1Q_5/2$. These two bounds do not agree with each other. This discrepency has to be sorted out.

We have listed these problems  for the sake of record. Our focus in this note is going to be yet another problem in the context of superstrings moving on $AdS_3$ backgrounds. In their study of  superstrings Giveon-Kutasov-Seiberg found a mysterious doubling of N=4 superconformal algebra. They provided rationale for this superalgebra doubling. This explanation is conjectural.

In the rest of this note we shall first of all formulate this problem and then discuss the solution provided by GKS. After that we shall present some details from our construction of an ADHM sigma model \cite{Ali:2023csc,Ali:2023icn}complementary to the one constructed by Witten in Ref. \cite{Witten:1994tz}. We then propose our resolution to the problem of superalgebra doubling in $AdS_3$ superstrings.

\section{The GKS Formalism}

Giveon-Kutasov-Seiberg revived the old fashioned worldsheet perturbative formalism to study propagation of strings and superstrings on $AdS_3 \times S^3 \times K3$ and $AdS_3 \times S^3 \times T^4$ backgrounds. They derived the corresponding spacetime conformal field theory, Virasoro and current algebras. It was an establishment of the $AdS_3/CFT_2$ conjecture. (Another derivation of the same was provided by Eberhardt-Gaberdiel-Gopakumar in \cite{Eberhardt:2019ywk} from a different angle.)

The bosonic string on $AdS_3$ times some compact manifold has the Lagrangian
\begin{equation}\label{cmlag}
    \mathcal{L}=\frac{2l^2}{l_{s}^2}(\partial \phi \Bar{\partial}\phi + e^{2\phi}\Bar{\partial}\gamma \partial \gamma)
    \end{equation}
where $l_s$ is the fundamental string length and $l$ is the radius of the curvature of the hyperspace used to define $AdS_3$ metric. A $B$-field of the form
\begin{equation}\label{bfield}
    B=l^2 e^{2\phi}d\gamma \wedge d \Bar{\gamma}
\end{equation}
has been incorporated and the $AdS_3$ metric is 
\begin{equation}\label{adsmetric}
    ds^2=l^2(d\phi^2 + e^{2\phi}\partial \gamma \partial \Bar{\gamma}).
\end{equation}
The theory has affine $SL(2,R) \times SL(2,R)$ Lie algebra symmetry at level $k$
\begin{equation}\label{lasym}
    J^{a}(z)J^{b}(z)=\frac{k\eta^{AB}/2}{(z-w)^2}+\frac{i\eta_{CD}\epsilon^{ABC}J^D(w)}{z-w}+...
\end{equation}
with $\eta^{AB}$ being the $SL(2,R)$ metric with signature $(+,+,-)$ and $\epsilon_{ABC}$ are the structure constants of $SL(2,R)$. Also the level $k$ is related to $l$ and $l_s$ by
\begin{equation}\label{llsk}
    l^2=l^2_{s}k.
\end{equation}
The central charge of the model is 
\begin{equation}\label{centchar}
    c=\frac{3k}{k-2}.
\end{equation}
The current algebra has a Wakimoto free field representation 
\begin{eqnarray}\label{wffr}
  J^-&=&\beta  ,~~~ J^3=\beta \gamma + \frac{\alpha_+}{2}\partial\phi, \nonumber\\
     J^+&=&\beta \gamma ^2+\alpha_+\gamma \partial \phi + k\partial \gamma   
\end{eqnarray}
with
\begin{equation}\label{alphap}
    \alpha_{+}^2=2k-4.
\end{equation}
The relevant vertex operators are
\begin{equation}\label{verop}
V_{jm\Bar{m}}=\gamma^{j+m}\Bar{\gamma}^{j+\Bar{m}}exp\left(\frac{2j}{{\alpha_+}}\phi\right)
\end{equation}
with scaling dimension of $V_{jm\Bar{m}}$ being
\begin{equation}\label{scadim}
    h=-\frac{j(j+1)}{k-2}.
\end{equation}
The unitarity condition puts a condition on the quantum numbers
\begin{equation}\label{unicond}
    -1 < j < \frac{k}{2}-1
\end{equation}
as analyzed in \cite{Dixon:1989cg} and \cite{Evans:1998qu}. The presence of above $SL(2,R)$ Lie algebra gives rise to three conserved charges
\begin{eqnarray}\label{threecc}
    L_0=-\oint dz J^3(z),~L_1&=&-\oint dz J^{+}(z),~
    L_{-1}=-\oint dz J^{-}(z)
\end{eqnarray}
which is generalized to full Virasoro algebra along the lines of Ref. \cite{Brown:1986nw} in the following way. The full Virasoro algebra can be introduced using affine Kac-Moody algebra in the following manner. If the string background is $AdS_3 \times \mathcal{N}$ with $\mathcal{N}$ having a worldsheet affine Lie algebra $\hat{G}$ for a compact group $G$ with currents $K^a(z)$ satisfying
\begin{equation}\label{kazcur}
  K^a(z)K^b(z)=  \frac{k'\delta^{ab}/2}{(z-w)^2}+\frac{if^{ab}_{~c}k'(w)}{(z-w)}+...,~a,b,c=1,...,\text{dimG}
\end{equation}
then the spacetime theory can be endowed with an affine Kac-Moody algebra with the generators
\begin{equation}\label{akmagen}
    T^a_n=\oint dz K^a(z) \gamma^n(z)
\end{equation}
and the algebra closes with the spacetime level 
\begin{equation}\label{stlevel}
    k_{s-t} \equiv \Tilde{k}=pk'
\end{equation}
with the integer
\begin{equation}\label{integer}
    p \equiv \oint dz \frac{\partial_{z}\gamma}{\gamma}.
\end{equation}

The Virasoro generators are defined as 
\begin{equation}\label{viragen}
    L_n=-\oint dz [(1-n^2)J^3\gamma^n+\frac{n(n-1)}{2}J^{-}\gamma^{n+1}+\frac{n(n+1)}{2}J^{+}\gamma^{n-1}].
\end{equation}
The corresponding spacetime Virasoro algebra has the central charge
\begin{equation}\label{cstkp}
    c_{s-t}=6kp.
\end{equation}
To discuss fermionic strings GKS introduced fermions $\psi^A(z)$ and $\chi^a(z)$ and `bosonic' currents $j^a(z)$ and $k^a(z)$ with
\begin{eqnarray}\label{ferbosc}
    J^A=j^A-\frac{i}{k}\epsilon^{A}_{~BC}\psi^B\psi^C,~
   K^a= k^a-\frac{i}{k'}f^{a}_{~bc}\chi^b\chi^c 
\end{eqnarray}
with two point functions
\begin{eqnarray}\label{twopfunc}
    \langle \psi^A(z)\psi^B(w)\rangle=\frac{k\eta^{AB}/2}{z-w},~
    \langle\chi^{a}(z)\chi^{b}(w)\rangle=\frac{k'\delta^{ab}/2}{z-w}.
\end{eqnarray}
With $A,B=1,2,3 ~\text{and}~ a,b=1,..,\text{dimG}$.

The worldsheet supercurrent $G(z)$ is proposed to be
\begin{equation}\label{wssupc}
    G(z)=\frac{2}{k}(\eta_{AB}\psi^Aj^B-\frac{i}{3k}\epsilon_{ABC}\psi^A\psi^B\psi^C)+\frac{2}{k'}(\chi^a k^a-\frac{i}{3k'}f_{abc}\chi^{a}\chi^{b}\chi^{c})+G_{rest}.
\end{equation}
When we take up the discussion of superstrings on $AdS_3 \times S^3 \times T^4$ then the worldsheet theory has a three sphere or $SU(2)$ WZW model. The $AdS_3$ currents are $(\psi^A,J^A)$ and $SU(2)$ currents are $(\chi^a,K^a)$ and the central charge on the worldsheet is
\begin{equation}\label{ccws}
    c=\left[\frac{3(k+2)}{k}+\frac{3}{2}\right]+\left[\frac{3(k'+2)}{k'}+\frac{3}{2}\right].
\end{equation}
The energy-momentum tensor $T(z)$ and the supercurrent $G(z)$ in this case are
\begin{equation}\label{emttz}
   T(z)=\frac{1}{k}(j^Aj_A-\psi^A\partial \psi_A)+\frac{1}{k}(k^ak_a-\chi^a\partial\chi_a)+\frac{1}{2}(\partial Y^i \partial Y_i-\lambda^i\partial\lambda_i)   
\end{equation}
   and
\begin{equation}\label{scgz}
     G(z)=\frac{2}{k}(\psi^Aj_A-\frac{i}{3k}\epsilon_{ABC}\psi^A\psi^B\psi^C)+\frac{2}{k}(\chi^a k^a-\frac{i}{3k}\epsilon_{abc}\chi^{a}\chi^{b}\chi^{c})+\lambda^i\partial\lambda_i.
\end{equation}
 Where $Y^i(z),~i=1,2,3,4$ are the four free bosons corresponding to $T^4$.

\section{Superalgebra Doubling}
With above GKS formalism we would now like to demand spacetime supersymmetry in the theory. It is well known that to  have N=1 spacetime supersymmetry the worldsheet superconformal symmetry must be enhanced to N=2.

The N=1 superconformal algebra has only two generators: the energy-momentum tensor $T(z)$ and super-tensor $G(z)$. Both of these have been constructed above. N=2 superconformal symmetry has four generators-the energy-momentum tensor $T(z)$, two super-tensors $G^{\pm}(z)$ and a $U(1)$ current $J(z)$.

GKS devised an ingenious method to get N=2 superconformal symmetry on worldsheet to get spactime supersymmetry. In particular they abandoned the usual approach of identifying the $U(1)$ currents. The reason is that this usual approach, discussed in their Appendix B, suffers from the problem of superalgebra doubling. We now spell out this problem. The N=2 $U(1)$ current is written as sum of parts corresponding to isometries
\begin{equation}\label{isomets}
    \frac{SL(2,R)}{U(1)} \times \frac{SU(2)}{U(1)} \times U(1)^2 \times U(1)^4
\end{equation}
\begin{equation}\label{jz1234}
    J(z)=J^{(1)}+J^{(2)}+J^{(3)}+J^{(4)}
\end{equation}
with
\begin{eqnarray}\label{jzful}
    J^{(1)}(z)&=&\frac{2}{k}(J^3+i\psi^1\psi^2),~
    J^{(2)}(z)=\frac{2}{k}(i\chi^1\chi^2-K^3) \nonumber\\
    J^{(3)}(z)&=&\frac{2}{k}\psi^3\chi^3,~~~~~~~~~~~~
    J^{(4)}(z)=i\lambda^1\lambda^2+i\lambda^3\lambda^4.
\end{eqnarray}
Here $J^3$ and $K^3$ are total $SL(2,R)$ and $SU(2)$ currents respectively. To construct spacetime supercharges $J(z)$ is bosonised as 
\begin{equation}\label{jzh123}
    J(z)=i\partial(H+H_1+H_2+H_3)
\end{equation}
with
\begin{eqnarray}\label{jzhfull}
i\partial H&=& \frac{2}{k}(-i\psi^1\psi^2+i\chi^1\chi^2+J^3-K^3),~
i\partial H_1= \frac{2}{k}\psi^3\chi^3,\nonumber \\
\partial H_2&=&\lambda^1\lambda^2, ~
\partial H_3=\lambda^3\lambda^4.
\end{eqnarray}
The spacetime charges are 
\begin{eqnarray}\label{stcq}
    Q_{\alpha}^{\pm}=\oint dz S^{\pm}_{\alpha},~\Bar{Q}_{\alpha}^{\pm}=\oint dz \Bar{S}^{\pm}_{\alpha}
\end{eqnarray}
with
\begin{eqnarray}\label{salpha}
    S_{\alpha}^{\pm}=e^{\pm \frac{i}{2}(H_2+H_3)},~S_{\Bar{\alpha}}^{\pm}=e^{\pm \frac{i}{2}(H_2-H_3)} 
\end{eqnarray}
Now the spacetime superalgebra is 
\begin{eqnarray}\label{stsal}
  \{ Q_{\alpha}^{+},Q_{\beta}^{-}\}&=&\delta_{\alpha\beta}(J^3-K^3),~
     \{\Bar{Q}_{\alpha}^{+},\Bar{Q}_{\beta}^{-}\}=\delta_{\Bar{\alpha}\Bar{\beta}}(J^3+K^3) \nonumber\\
     \{ Q_{\alpha}^{-},\Bar{Q}_{\beta}^{+}\}&=&\{ Q_{\alpha}^{+},\Bar{Q}_{\Bar{\beta}}^{-}\} = \gamma^{i}_{\alpha\Bar{\beta}}P_i.
\end{eqnarray}
First of these equations is the Ramond sector N=4 superalgebra. With that interpretation the second of the above equations give another copy of the Ramond sector superalgebra.

For a giveN theory there should be only one (small) N=4  superconformal algebra. Here we have two copies. Why do we have two copies of the Ramond sector (small)  N=4  superconformal algebra is the problem we are trying to resolve in this note.

\section{The GKS Resolution}
One immediate idea to interpret this superalgebra doubling is to propose that the second equation of (\ref{stsal}) is the copy of superalgebra corresponding to the other chirality. GKS reject this possibility on the basis of two rather stark arguments. Firstly the left and right moving generators commute or anti-commute, that is, have zero commutators or anti-commutators with the other chirality. From third of Eqn.(\ref{stsal}) it is clear that the superalgebra of two copies of the superalgebra have non-zero anti-commutators. Secondly the worldsheet and spacetime chiralities in string theory are related. Thus a single worldsheet chirality giving rise to both  chiralities in the spacetime theory is not feasible.  Moreover the other chirality of worldsheet too will give rise to both chiralities on the spacetime side.

The GKS solution to the Ramond superalgebra doubling is to propose that the two superalgebras are different Ramond twists \cite{Schwimmer:1986mf} of the NS generators
\begin{eqnarray}\label{nsgen}
    T^3_n(\eta)&=&T^3_n(0)-\frac{\eta kp}{2}\delta_{n,0},~
     T^{\pm}_{n\pm \eta}=T^{\pm}_n(0), \nonumber \\
    Q^{1}_{n+ \frac{\eta}{2}}(\eta)&=&Q^{1}_{\eta}(0),~~ Q^{2}_{n- \frac{\eta}{2}}(\eta)=Q^{2}_{n}(0), \nonumber \\  
      L_n(\eta)&=&L_n(0)-\eta  T^3_n(0)+ \eta^2\frac{kp}{4}\delta_{n,0}.
\end{eqnarray}
Now since both vacua cannot be described at the same time they proposed the following procedure to restrict to single sector of the vacuum. If $J(z)$ contains the combination $J^3-K^3$ then the physical states are defined by restricting to the cohomology of operators $Q_{\alpha}^{\pm}$:
\begin{eqnarray}\label{cohop}
    Q_{\alpha}^{\pm} |phy\rangle=0,~|phy\rangle \sim |phy\rangle +Q|anything\rangle.
\end{eqnarray}
Because of (\ref{stsal}) such states have $J^3-K^3=0$. After that the four remaining supercharges $ \Bar{Q}_{\Bar{\alpha}}^{\pm}$ together with $J^3+K^3$ form the left moving small N=4 superconformal algebra 
\begin{eqnarray}\label{small1}
\left[{L}_m, {L}_n\right]& = &
(m-n){L}_{m+n}+\frac{{c}}{12}
(m^3-m)\delta_{{m+n},0},\nonumber\\
\left[{L}_m, {G}_r^a\right]& = &
\left(\frac{m}{2}-r\right){G}^a_{m+r},
~~~\left[{L}_m, {J}^i_n\right]=-n{J}^i_{m+n},\nonumber\\
\{{G}^a_r, {G}^b_s\}& = &2\delta^{ab}{L}_{r+s}
-4(r-s)\alpha^{+i}_{ab}{J}^i_{r+s}+\frac{{c}}{3}
\left(r^2-\frac{1}{4}\right)\delta^{ab}\delta_{{r+s},0},\nonumber\\
\left[{J}^i_m, {G}^a_r\right]& = &
\alpha^{+i}_{ab}{G}^b_{m+r},
~~~~\left[{J}^i_m, {J}^j_n\right]
=\epsilon^{ijk}{J}^k_{m+n}-m\frac{{k}}{2}\delta^{ij}
\delta_{m+n,0},\nonumber\\
\alpha^{+i}_{ab}& = &
\frac{1}{2}\left(\delta^i_a\delta^0_b-\delta^i_b\delta^0_a\right)
+\frac{1}{2}\epsilon^{iab}, ~~~{c}=6{k}.
\end{eqnarray}
When the current $J(z)$ contains $J^3+K^3$ the roles of $Q$ and $\Bar{Q}$ as well as $J^3-K^3$ and $J^3+K^3$ are reversed.

By this procedure the states of the other chirality are killed. Moreover, GKS argue, that this procedure ensures $P^i=0$ using the unitarity argument. The projecting out of the states of other chirality by Eqn.(\ref{cohop}) the worldsheet and spactime chirality correlation is restored. They also compare the projection (\ref{cohop}) to the topological twist \cite{Eguchi:1990vz,Ali:1991tj,Nojiri:1991fn} of N=2 theory whose physical states are in one to one correspondence  with the chiral ring of the original theory.

This resolution of the superalgebra doubling is admittedly conjectural and kills half of the spectrum in an ad hoc manner.

\section{Our Resolution}

To present our resolution to the problem of doubling of  Ramond superalgebra we need some background from Refs.\cite{Ali:2023csc,Ali:2023icn,Witten:1994tz} (see also Ref. \cite{Ali:2023zxt}). In \cite{Witten:1994tz} an ADHM  instanton sigma model was constructed whose action is
\begin{equation}\label{action1}
    S_W=S^{kin}+S^{int}
\end{equation}
with 
\begin{eqnarray}\label{action2}
 S^{kin} &=& \int d^2 \sigma ( \epsilon_{AB} \epsilon_{YZ} \partial_{-}  X^{AY}\partial_{+} X^{BZ}+ i\epsilon_{A^{\prime}B^{\prime}}\epsilon_{YZ}\psi_-^{A^{\prime}Y} \partial_{+} \psi_-^{B^{\prime}Z}  \nonumber \\
&+&\epsilon_{A^{\prime}B^{\prime}}\epsilon_{Y^{\prime}Z^{\prime}}\partial_{-}\phi^{A^{\prime}Y^{\prime}}\partial_{+} \phi^{B^{\prime}Z^{\prime}}
+i\epsilon_{AB}\epsilon_{Y^{\prime}Z^{\prime}}\chi_-^{AY^{\prime}}\partial_{+} \chi_{-}^{BZ^{\prime}} \nonumber \\&+&i\lambda_{+}^{a} \partial_{-} \lambda_+^{a})
\end{eqnarray}
and
\begin{eqnarray}\label{action3}
    S^{int}&=&-\frac{i}{2}m\int d^2 \sigma\lambda^{a}_{+}[(\epsilon^{BD} \frac{\partial C^{a}_{BB'}}{\partial X^{DY}}\psi^{B'Y}_{-} + \epsilon^{B'D'}\frac{\partial C^{a}_{BB'}}{\partial \phi^{D'Y'}}{\chi}^{BY'}_{-})\nonumber \\&-& \frac{im}{4} \epsilon^{AB}\epsilon^{A'B'}C^a_{AA'}C^a_{BB'})]
\end{eqnarray}
It has $4k$ bosons $X^{AY}$, $A = 1,2,$ $Y = 1,2,...,2k$ and $4k'$ bosons  $\phi^{A'Y'}$, $A' = 1,2,$ $Y'= 1,2,...,2k'$ and corresponding right-handed superpartners $\psi_{-}^{A'Y}$ for  $X^{AY}$, and $\chi_{-}^{AY'}$ for $\phi^{A'Y'}$. There
are also left-handed fermions $\lambda_{+}^{a}$, $a=1,2,...,n$. The indices  are
raised and lowered by taking the help of antisymmetric tensors $\epsilon^{AB} (\epsilon_{AB})$, $\epsilon^{A'B'} (\epsilon_{A'B'})$, $\epsilon^{YZ} (\epsilon_{YZ})$ and $\epsilon^{Y'Z'} (\epsilon_{Y'Z'})$ belonging to $SU(2)$, $SU(2)'$, $Sp(2k)$ and
$Sp(2k')$ respectively.  The conditions on $C^{a}_{AA'}$ are
\begin{eqnarray}\label{adhm1}
\frac{\partial C_{AA'}^a}{\partial X^{B Y}}
+\frac{\partial C_{BA'}^a}{\partial X^{A Y}}
=0=  \frac{\partial C_{AA'}^a}{\partial \phi^{B'Y'}}
+\frac{\partial C_{AB'}^a}{\partial \phi^{A'Y'}},
\end{eqnarray}

\begin{equation}\label{adhm2}
\sum_{a}(C^{a}_{AA'}{C}^{a}_{BB'}+{C}^{a}_{BA'}{C}^{a}_{AB'})=0.
\end{equation}
In Ref.\cite{Ali:2023csc} we constructed a Complementary ADHM instanton sigma model with the action

\begin{equation}\label{action4}
 S_C=\hat S^{kin}+\hat S^{int}
\end{equation}
with 
\begin{eqnarray}\label{action5}
 \hat S^{kin} &=& \int d^2 \sigma ( \epsilon_{AB} \epsilon_{YZ} \partial_{-}  X^{AY}\partial_{+} X^{BZ}+ i\epsilon_{A^{\prime}B^{\prime}}\epsilon_{YZ}\psi_-^{A^{\prime}Y} \partial_{+} \psi_-^{B^{\prime}Z}  \nonumber \\
&+&\epsilon_{A^{\prime}B^{\prime}}\epsilon_{Y^{\prime}Z^{\prime}}\partial_{-}\phi^{A^{\prime}B^{\prime}}\partial_{+} \phi^{B^{\prime}Z^{\prime}}
+i\epsilon_{AB}\epsilon_{Y^{\prime}Z^{\prime}}\chi_-^{AY^{\prime}}\partial_{+} \chi_{-}^{BZ^{\prime}} \nonumber \\&+&i\hat \lambda_+^{a^\prime} \partial_{-}\hat \lambda_+^{a^\prime})
\end{eqnarray}
and
\begin{eqnarray}\label{action6}
    \hat S^{int}&=&-\frac{i}{2}m\int d^2 \sigma\hat\lambda^{a^{\prime}}_{+}[(\epsilon^{BD} \frac{\partial \hat C^{a^{\prime}}_{BB'}}{\partial X^{DY}}\psi^{B'Y}_{-} + \epsilon^{B'D'}\frac{\partial \hat C^{a'}_{BB'}}{\partial \phi^{D'Y'}}{\chi}^{BY'}_{-})\nonumber \\&-& \frac{im}{4} \epsilon^{AB}\epsilon^{A'B'}\hat C^{a'}_{AA'}\hat C^{a'}_{BB'})]
\end{eqnarray}

with the conditions
\begin{eqnarray}\label{adhm3}
\frac{\partial \hat C_{AA'}^{a'}}{\partial X^{B Y}}
+\frac{\partial \hat C_{BA'}^{a'}}{\partial X^{A Y}}
=0=  \frac{\partial \hat C_{AA'}^{a'}}{\partial \phi^{B'Y'}}
+\frac{\partial \hat C_{AB'}^{a'}}{\partial \phi^{A'Y'}},
\end{eqnarray}

\begin{equation}\label{adhm4}
\sum_{a'}(\hat C^{a'}_{AA'}{\hat C}^{a'}_{BB'}+{\hat C}^{a'}_{BA'}{\hat C}^{a'}_{AB'})=0
\end{equation}
The tensors $C^{a}_{AA'}$ and $\hat C^{a'}_{AA'}$ are linear in both $X$ and $\phi$ and their general expression are
\begin{equation}\label{generalform1}
C^{a}_{AA'}=M^{a}_{AA'}+\epsilon_{AB} N^{a}_{A'Y}X^{BY}+\epsilon_{A'B'}D^{a}_{AY'}\phi^{~B'Y'}+\epsilon_{AB}\epsilon_{A'B'}E^{a}_{YY'}X^{BY}\phi^{~B'Y'}
\end{equation}
and
\begin{equation}\label{generalform2}
\hat C^{a'}_{AA'}=\hat M^{a'}_{AA'}+\epsilon_{AB} \hat N^{a'}_{A'Y}X^{BY}+\epsilon_{A'B'}\hat D^{a'}_{AY'}\phi^{~B'Y'}+\epsilon_{AB}\epsilon_{A'B'}\hat E^{a'}_{YY'}X^{BY}\phi^{~B'Y'}.
\end{equation}
For the one instanton cases the potentials in action (\ref{action1}) and \ref{action4}
become
\begin{equation}\label{potential}
V_W = \frac{m^2}{8}(X^{2}+\rho^{2})\phi^{2}~\text{and}~V_C = \frac{m^2}{8}(\phi^{2}+\omega^{2})X^{2}.
\end{equation}
These two models are independent of each other and there is a duality between them given by the following interchanges $X  \leftrightarrow \phi,~~k  \leftrightarrow k', ~~\psi \leftrightarrow \chi$.

We assert that the two copies of the Ramond superalgebra that Giveon-Kutasov-Seiberg get in their construction are the two distinct and independent branches of the theory just like the two branches of the moduli space of the ADHM instanton sigma model. Thus these should be treated separately. In particular there is no need to kill the Hilbert space of physical states of one branch or the other.

\section{Discussion}
 Above resolution comes with several questions. For example  why did we not discover the other branch of the moduli space so far? 
 Here we would like recall that as long back as 1988 Sevrin, Troost and van Proeyen \cite{Sevrin:1988ab} told us  that the small N=4 superconformal algebra is contained in large N=4 superconformal algebra in two different ways. They also pointed out that there is a duality between these two embeddings. Thus the doubling of small N=4 superconformal algebra is known for a long time.

 Another question concerns whether the lessons learned from ADHM instanton sigma models can be applied to $AdS_3$ superstrings or not. In this regard it is a long held belief that the ADHM instanton sigma model with (small) N=4 supersymmetry and flow to a superconformal limit in the infrared \cite{Witten:1995zh}.

Next it is clear from our construction of the Complete ADHM instanton sigma model \cite{Ali:2023icn} that the duality between the original and Complementary models become a $Z_2$ symmetry in case of the Complete model. This $Z_2$ symmetry plays a role in getting the complete free field realizations of the large N=4 superconformal symmetry \cite{Ali:2023kkf} that is relevant for  $AdS_3 \times S^3 \times S^3 \times S^1$ superstrings \cite{Gukov:2004ym}. In Refs. \cite{Ali:2025jcu, Ali:2025ntc} off-shell formalism for the complementary ADHM instanton sigma model was discussed.

Integrability of $AdS_3$ superstrings  \cite{Chakraborty:2022iuk, Sundin:2012gc, Dei:2018jyj} and the BMN dynamics\cite{Rughoonauth:2012qd} too get related to the issue investigated in the present not.

Finally we would like to point out a related develop in which in Ref. \cite{Papadopoulos:2024uvi} Papdopoulos and Witten gave a direct proof of the fact that in two dimensions scale invariance implies conformal invariance. In Ref. \cite{Ali:2024amc} the symmetry structure of ADHM instanton sigma models, $AdS_3$ superstrings and the $N=4$ superconformal algebras was compared.

 \textbf{Acknowledgements}: These investigations were made possible by the academic environment created by my students Dr. P.P. Abdul Salih, Dr. Shafeeq Rahman Thottoli and Dr. Mohsin Ilahi. I thank Prof. D.P. Jatkar and the Harish-Chandra Research Institute, Prayagraj, where part of this work was done, for hospitality. I thank Prof. Shahid Husain  and Prof. Isar A. Rizvi for moral support at various junctures.


\begin{thebibliography}{9}
\bibitem{Maldacena:1997re}
J.~M.~Maldacena,
``The Large N limit of superconformal field theories and supergravity,''
Adv. Theor. Math. Phys. \textbf{2} (1998), 231-252
[arXiv:hep-th/9711200 [hep-th]].

\bibitem{Witten:1998qj}
E.~Witten,
``Anti-de Sitter space and holography,''
Adv. Theor. Math. Phys. \textbf{2}, 253-291 (1998)
[arXiv:hep-th/9802150 [hep-th]].


\bibitem{Gubser:1998bc}
S.~S.~Gubser, I.~R.~Klebanov and A.~M.~Polyakov,
``Gauge theory correlators from noncritical string theory,''
Phys. Lett. B \textbf{428}, 105-114 (1998)
doi:10.1016/S0370-2693(98)00377-3
[arXiv:hep-th/9802109 [hep-th]].
	
	\bibitem{Giveon:1998ns}
	A.~Giveon, D.~Kutasov and N.~Seiberg,
	``Comments on string theory on AdS(3),''
	Adv.\ Theor.\ Math.\ Phys.\  {\bf 2} (1998) 733
	[hep-th/9806194].

	\bibitem{Elitzur:1998mm}
	S.~Elitzur, O.~Feinerman, A.~Giveon and D.~Tsabar,
	``String theory on $AdS_3 \times S^3 \times S^3 \times  S^1$,''
	Phys.\ Lett.\ B {\bf 449} (1999) 180
	[hep-th/9811245].
	
	\bibitem{Ali:2000we}
	A.~Ali,
	``Conformal symmetry of superstrings on $AdS_3 \times S^3 \times T^4$ and D1 / D5 system,''
	Mod.\ Phys.\ Lett.\ A {\bf 17} (2002) 2477
	[hep-th/0007021].
	
	


		\bibitem{Gukov:2004ym}
		S.~Gukov, E.~Martinec, G.~W.~Moore and A.~Strominger,
		``The Search for a holographic dual to $AdS_3 \times S^3 \times S^3 \times S^1$,''
		Adv.\ Theor.\ Math.\ Phys.\  {\bf 9} (2005) 435
		[hep-th/0403090].
		
		\bibitem{Tong:2014yna}
		D.~Tong,
		``The holographic dual of $AdS_{3} \times  S^{3} \times S^{3} \times S^{1}$,''
		JHEP {\bf 1404} (2014) 193
		[arXiv:1402.5135 [hep-th]].
		
		
		\bibitem{Gaberdiel:2010pz}
		M.~R.~Gaberdiel and R.~Gopakumar,
		``An $AdS_3$ Dual for Minimal Model CFTs,''
		Phys.\ Rev.\ D {\bf 83} (2011) 066007
		[arXiv:1011.2986 [hep-th]].
		
		\bibitem{Eberhardt:2017pty}
		L.~Eberhardt, M.~R.~Gaberdiel and W.~Li,
		``A holographic dual for string theory on $AdS_3 \times S^3 \times S^3 \times S^1$,''
		JHEP {\bf 1708} (2017) 111
		[arXiv:1707.02705 [hep-th]].
		
		\bibitem{Eberhardt:2017fsi}
		L.~Eberhardt, M.~R.~Gaberdiel, R.~Gopakumar and W.~Li,
		``BPS spectrum on $AdS_3 \times S^3 \times S^3 \times S^1$,''
		JHEP {\bf 1703} (2017) 124
		[arXiv:1701.03552 [hep-th]].
\bibitem{Sevrin:1988ab}
A. Sevrin, W. Troost and A. van Proeyen, ``Superconformal Algebras in Two-Dimensions with N=4'', Phys. Lett. B 208 (1988) 447.
  

\bibitem{Ali:2000zu}
A.~Ali,
``Free field realizations of N=4 superconformal algebras,''
Indian J. Pure Appl. Phys. \textbf{38}, 446-452 (2000)

\bibitem{Ali:2003aa}
A.~Ali,
``Types of two-dimensional N = 4 superconformal field theories,''
Pramana \textbf{61}, 1065-1078 (2003)
[arXiv:hep-th/9906096 [hep-th]].

\bibitem{Ivanov:1988rt}
	E.~A.~Ivanov, S.~O.~Krivonos and V.~M.~Leviant,
	``Quantum N=3, N=4 Superconformal WZW Sigma Models,''
	Phys.\ Lett.\ B {\bf 215} (1988) 689
	Erratum: [Phys.\ Lett.\ B {\bf 221} (1989) 432]

\bibitem{Banados:1992wn}
M.~Banados, C.~Teitelboim and J.~Zanelli,
Phys. Rev. Lett. \textbf{69}, 1849-1851 (1992)
doi:10.1103/PhysRevLett.69.1849
[arXiv:hep-th/9204099].

\bibitem{Ali:1992mj}
A.~Ali and A.~Kumar,
``O (d, d) transformations and 3-D black hole,''
Mod. Phys. Lett. A \textbf{8}, 2045-2052 (1993)
[arXiv:hep-th/9303032].

\bibitem{Sommovigo:2003kd}
L.~Sommovigo,
``Penrose limit of ${\rm AdS}_3 \times {\rm S}^3 \times {\rm S}^3 \times {\rm S}^1$ and its associated sigma model,''
JHEP \textbf{07}, 035 (2003)
[arXiv:hep-th/0305151 ].

\bibitem{Dei:2018yth}
A.~Dei, M.~R.~Gaberdiel and A.~Sfondrini,
``The plane-wave limit of ${\rm AdS}_3 \times {\rm S}^3 \times {\rm S}^3 \times {\rm S}^1$,''
JHEP \textbf{08}, 097 (2018)
[arXiv:1805.09154 [hep-th]].

\bibitem{Ali:1993sd}
A.~Ali and A.~Kumar,
``A New N=4 superconformal algebra,''
Mod. Phys. Lett. A \textbf{8}, 1527-1532 (1993)
[arXiv:hep-th/9301010].

\bibitem{Seiberg:1999xz}
N.~Seiberg and E.~Witten,
``The D1 / D5 system and singular CFT,''
JHEP \textbf{04}, 017 (1999)
[arXiv:hep-th/9903224 [hep-th]].

\bibitem{Strominger:1996sh}
A.~Strominger and C.~Vafa,
``Microscopic origin of the Bekenstein-Hawking entropy,''
Phys. Lett. B \textbf{379}, 99-104 (1996)
[arXiv:hep-th/9601029 [hep-th]].



	\bibitem{Ali:2023csc}
	A.~Ali and M.~Ilahi,
	``Complementary ADHM Instanton Sigma Model,''
	[arXiv:2305.05951 [hep-th]].

\bibitem{Ali:2023icn}
A.~Ali and P.~P.~A.~Salih,
``Complete ADHM Sigma Model,''
[arXiv:2305.09516 [hep-th]].


		\bibitem{Witten:1994tz}
		E.~Witten,
		``Sigma models and the ADHM construction of instantons,''
		J. Geom. Phys. \textbf{15} (1995), 215-226
		[arXiv:hep-th/9410052 [hep-th]].


\bibitem{Eberhardt:2019ywk}
L.~Eberhardt, M.~R.~Gaberdiel and R.~Gopakumar,
``Deriving the AdS$_{3}$/CFT$_{2}$ correspondence,''
JHEP \textbf{02}, 136 (2020)
[arXiv:1911.00378 [hep-th]].

\bibitem{Dixon:1989cg}
L.~J.~Dixon, M.~E.~Peskin and J.~D.~Lykken,
``N=2 Superconformal Symmetry and SO(2,1) Current Algebra,''
Nucl. Phys. B \textbf{325}, 329-355 (1989)

\bibitem{Evans:1998qu}
J.~M.~Evans, M.~R.~Gaberdiel and M.~J.~Perry,
``The no ghost theorem for AdS(3) and the stringy exclusion principle,''
Nucl. Phys. B \textbf{535}, 152-170 (1998)
[arXiv:hep-th/9806024 [hep-th]].
  
\bibitem{Brown:1986nw}
J.~D.~Brown and M.~Henneaux,
``Central Charges in the Canonical Realization of Asymptotic Symmetries: An Example from Three-Dimensional Gravity,''
Commun. Math. Phys. \textbf{104}, 207-226 (1986)


\bibitem{Schwimmer:1986mf}
A.~Schwimmer and N.~Seiberg,
``Comments on the N=2, N=3, N=4 Superconformal Algebras in Two-Dimensions,''
Phys. Lett. B \textbf{184}, 191-196 (1987)

\bibitem{Eguchi:1990vz}
T.~Eguchi and S.~K.~Yang,
``N=2 superconformal models as topological field theories,''
Mod. Phys. Lett. A \textbf{5}, 1693-1701 (1990)



\bibitem{Ali:1991tj}
A.~Ali, D.~P.~Jatkar and A.~Kumar,
``Algebraic structure of a topological superconformal theory,''
IP-BBSR-91-14.
  
\bibitem{Nojiri:1991fn}
S.~Nojiri,
``N=2 superconformal topological field theory,''
Phys. Lett. B \textbf{264}, 57-61 (1991)

\bibitem{Ali:2023zxt}
A.~Ali, M.~Ilahi and S.~R.~Thottoli,
``Quantization of Complementary ADHM Sigma Model,''
[arXiv:2306.10002 [hep-th]].

\bibitem{Witten:1995zh}
E.~Witten,
``Some comments on string dynamics,''
[arXiv:hep-th/9507121 [hep-th]].


\bibitem{Ali:2023kkf}
A.~Ali and M.~Ilahi,
``$Z_2$ Symmetry of $AdS_3 \times S^3 \times S^3 \times S^1$ Superstrings,''
[arXiv:2306.13970 [hep-th]].

\bibitem{Ali:2025ntc}
A.~Ali, M.~Ilahi, P.~P.~A.~Salih and S.~R.~Thottoli,
``Harmonic Superspace for Ali-Ilahi's ADHM Instanton Sigma Model,''
[arXiv:2507.22948 [hep-th]].

\bibitem{Ali:2025jcu}
A.~Ali, M.~Ilahi, P.~P.~A.~Salih and S.~R.~Thottoli,
``Off-shell Formalism for Ali-Ilahi's ADHM Instanton Sigma Model,''
[arXiv:2507.11305 [hep-th]].

\bibitem{Chakraborty:2022iuk}
A.~Chakraborty, R.~R.~Nayak, P.~Pandit and K.~L.~Panigrahi,
``Neumann-Rosochatius system for rotating strings in AdS$_{3}$ \texttimes{} S$^{3}$ \texttimes{} S$^{3}$ \texttimes{} S$^{1}$ with flux,''
JHEP \textbf{12}, 059 (2022)
[arXiv:2209.07379 [hep-th]].

\bibitem{Sundin:2012gc}
P.~Sundin and L.~Wulff,
``Classical integrability and quantum aspects of the AdS(3) x S(3) x S(3) x S(1) superstring,''
JHEP \textbf{10}, 109 (2012)
[arXiv:1207.5531 [hep-th]].

\bibitem{Dei:2018jyj}
A.~Dei and A.~Sfondrini,
``Integrable S matrix, mirror TBA and spectrum for the stringy AdS$_{3}$ \texttimes{} S$^{3}$ \texttimes{} S$^{3}$ \texttimes{} S$^{1}$ WZW model,''
JHEP \textbf{02}, 072 (2019)
[arXiv:1812.08195 [hep-th]].

\bibitem{Rughoonauth:2012qd}
N.~Rughoonauth, P.~Sundin and L.~Wulff,
``Near BMN dynamics of the AdS(3) x S(3) x S(3) x S(1) superstring,''
JHEP \textbf{07}, 159 (2012)
[arXiv:1204.4742 [hep-th]].

\bibitem{Papadopoulos:2024uvi}
G.~Papadopoulos and E.~Witten,
``Scale and Conformal Invariance in 2d Sigma Models, with an Application to N=4 Supersymmetry,''
[arXiv:2404.19526 [hep-th]].

\bibitem{Ali:2024amc}
A.~Ali, M.~Ilahi, P.~P.~A.~Salih and S.~R.~Thottoli,
``Symmetry Structure of ADHM Sigma Models and $AdS_3$ Superstrings,''
[arXiv:2409.00474 [hep-th]].

\end{thebibliography}
\end{document}